\documentclass[manuscript]{emulateapj-rtx4}
\pdfoutput=1
\usepackage{natbib,graphicx,amsmath}
\newcommand{\parz}[1]{\frac{\partial #1}{\partial z}}

\newcommand{\pard}[2]{\frac{\partial #1}{\partial #2}}

\begin{document}

\title{Tidal Evolution of Close-in Extrasolar Planets: High Stellar Q from New Theoretical Models}

\author{Kaloyan Penev}
\author{Dimitar Sasselov}
\affil{60 Garden St., M.S. 16, Cambridge, MA 02138}

\begin{abstract}
In recent years it has been shown that the tidal coupling between extrasolar
planets and their stars could be an important mechanism leading to orbital
evolution. Both the tides the planet raises on the star and vice versa are
important and dissipation efficiencies ranging over four orders of magnitude
are being used. In addition, the discovery of extrasolar planets extremely
close to their stars has made it clear that the estimates of the tidal
quality factor, Q, of the stars based on Jupiter and its satellite system and
on main sequence binary star observations are too low, resulting in lifetimes
for the closest planets orders of magnitude smaller than their age.  We argue
that those estimates of the tidal dissipation efficiency are not applicable
for stars with spin periods much longer than the extrasolar planets' orbital
period. We address the problem by applying our own values for the dissipation
efficiency of tides, based on our numerical simulations of externally
perturbed volumes of stellar-like convection. The range of dissipation we
find for main-sequence stars corresponds to stellar $Q_*$ of $10^8$ to
$3{\times}10^9$. The derived orbit lifetimes are comparable to, or much
longer than the ages of the  observed extrasolar planetary systems. The
predicted orbital decay transit timing variations due to the tidal coupling
are below the rate of ms/yr for currently known systems, but within reach of
an extended Kepler mission provided such objects are found in its field.
\end{abstract}

\subjectheadings{
	Convection; Turbulence; Methods: numerical; Planets and satellites:
	dynamical evolution and stability; Planet-star interactions; planetary systems}

\section{Introduction}
\label{sec: intro}
With the discovery of several giant extrasolar planets that are extremely
close to their
stars, it has become clear that the dissipation of the tides that the
planet raises on the star plays an important role in determining the orbital
evolution \citep[c.f.][among others]{Dobbs-Dixon_et_al_04, Ibgui_Burrows_09,
Pont_09, Jackson_et_al_09b, Jackson_et_al_08d, Jackson_et_al_08a,
Matsumura_Takeda_Rasio_08}. The problem already has a history
\citep[e.g.][]{Rasio_et_al_96, Terquem_et_al_98, Sasselov_03,
Ogilvie_Lin_04},
which led to the basic conclusion that a comprehensive theoretical
understanding of turbulent dissipation in stellar convection zones is
lacking. The recent discovery of the the very close-in giant planet WASP-19b
made that very clear \citep{Hebb_et_al_10}.

In some cases the dissipation of the tides on the
planet is important as well, but (at least for single planet systems) this
eventually leads to the planet being on a
circular orbit with its spin synchronized with the star, at which point the
tides on the planet become independent of time and hence no dissipation occurs.
The tides raised on the star will also have the tendency to synchronize the
rotation of the star with the orbit, but in the case of planets usually the
angular momentum of the planetary orbit is not enough to achieve that. 

Commonly the rate at which energy is dissipated from the stellar tides is
parametrized in terms of a quality factor ($Q_*$) defined as the fraction of
the tidal energy dissipated in one tidal period. It is then assumed that this
value depends only on the star, and not on the frequency of the tides and its
value derived from observations of main sequence solar mass binary stars
\citep{Meibom_Mathieu_05} is used: $10^5\lesssim Q_* \lesssim 10^7$. This
dissipation, however, is found to lead to unreasonably short lifetimes for at
least two systems --- Wasp 18b \citep{wasp18_nature} and WASP-19b
\citep{Hebb_et_al_10}. For these systems the time it would take for the
planets to plunge into the star, if the above $Q_*$ values are assumed, is
about three orders of magnitude smaller than the estimated ages of the
systems and four orders of magnitude smaller than the total lifetime of the
star, making it extremely unlikely that such a system should ever have been
observed \citep{wasp18_nature}, unless one assumes a mechanism that
continuously resupplies these extremely close to the star regions with fresh
planets.

There is however a fundamental difference between the tides in a binary star
and a star--planet system, namely that the mass of the secondary in the case
of stars is large enough to spin the primary to synchronous rotation with the
orbit, and in the case of planets it is not. And indeed for all transiting
planetary systems for which the spin period of the star ($p_*$) is known it
is found to be much longer than the orbital period ($P$). This distinction is
important, because when $|P|>|p_*/2|$ the time variable tidal perturbation
can resonantly excite inertial waves in the star, thus resulting in several
orders of magnitude larger shear, and hence dissipation, compared to the
static tide \citep{Savonije_Papaloizou_97, Dintrans_Ouyed_01, Ogilvie_Lin_04,
Wu_05a, Wu_05b, Papaloizou_Ivanov_05, Ogilvie_Lin_07, Ivanov_Papaloizou_07,
Ogilvie_09}.

For stars with surface convective zones the turbulent convective motions are
thought to be causing a cascade of energy to occur from the large scales of the
tides to smaller and smaller scales until eventually the finite viscosity of the
plasma becomes important and converts this energy into heat. The usual treatment
of this complicated process is to assume that it behaves like an effective
turbulent viscosity coefficient \citep[c.f.][]{Zahn_66, Goldreich_Nicholson_77}. 
However, the different prescriptions have been difficult to reconcile with each
other and with different observations \citep{Goodman_Oh_97}. 

\citet{our_direct_viscosity, our_code, our_k_dwarfs, our_sun_perturbative} used
numerical simulations of
stellar--like convection to show that the dissipative properties of the
turbulent convective flow are indeed well approximated by an effective
viscosity coefficient. Therefore the simulations allow us to derive its
theoretical value.  In this paper we combine these new results to construct a
complete prescription for the effective viscosity and show that the resulting
dissipation, in the absence of resonantly excited tidal waves, produces much
higher $Q_*$ values than those found observationally for main sequence stars,
and that those values are consistent with even the strongest current
constraints on the tidal dissipation efficiency. In addition we derive the
magnitude of the expected transit timing variations (TTV) due to orbital
decay and compare them to possible observational contraints (e.g. by the
Kepler mission) that could provide a direct test of the proposed viscosity
prescription.

The organization of the paper is as follows: in Section \ref{sec: classical}
we introduce what is usually understood by the $Q_*$ parameter, in Section
\ref{sec: viscosity} we discuss how we arrive at the turbulent viscosity we
use to calculate orbital decay, in Section \ref{sec: torque} we follow
\citet{Scharlemann_81, Scharlemann_82} to convert our turbulent viscosity to
a tidal torque, in Section \ref{sec: stellar models} we present the stellar
structure and evolution models used in the calculation of the torque, in
Section \ref{sec: orbit} we calculate the effective $Q_*$ which corresponds
to this torque, the TTVs that our estimate of the dissipation would result in
and the future lifetimes of close in extrasolar planets, finally in Section
\ref{sec: discussion} we summarize our results.

\section{Classical Approach to Q}
\label{sec: classical}
The tide raised by a planet on its star is a quadrupole wave of amplitude $h$; the tidal
motions are of order $({\omega}-{\Omega})h$ which set up shear inside the star. The 
orbital angular frequency, $\omega$, exceeds the stellar spin one, $\Omega$, in the case
we consider here, and the tidal forcing due to $\omega$ occurs with timescales that are
much shorter than the correlation time of the turbulence in the star's convection zone.

Dissipation in the star causes the tidal bulge $h$ to lag behind by an angle ${\delta}$,
which is determined by the amount of coupling between the tide and the source of the
dissipation, presumably the turbulent eddies in the convection zone. One could compare
the response of the star to that of a forced harmonic oscillator and relate ${\delta}$ to
a specific dissipation function $Q = ({\omega}-{\Omega})E_0/{\dot{E}} = 1/2\delta$ 
\citep{Murray_Dermott_99}. This is how the tidal dissipation quality factor $Q_*$ is defined,
with $E_0$ being the energy stored in the tidal bulge, and
$\dot{E}$, the rate of viscous dissipation of energy. This is another way to determine
the lag angle ${\delta}$. 

The turbulent 
viscosity is introduced in the sense of Rayleigh-Benard incompressible
convection, ${\nu}_{\rm t}={\frac{1}{3}}vl\approx l^2/{{\tau}}$, for a 
low forcing frequency. Here, $l$ is the convective mixing length, $v$ is
the convective velocity, and ${\tau}$ is the convective timescale
(eddy turnover time). Then basically $Q_*\propto GM/r_*{\omega}{\nu}_{\rm t}$, where
$M$ and $r_*$ are the stellar mass and radius \citep[see][for more
details]{Sasselov_03}.
 
\section{Effective Turbulent Viscosity}
\label{sec: viscosity}
We need to combine the perturbatively derived effective viscosity of
\citet{our_k_dwarfs} based on realistic low mass star convective models with
the \citet{our_direct_viscosity} direct viscosity from simulations with
external forcing in order to get a complete and reliable prescription for the
full viscosity tensor.

As discussed in \citet{our_k_dwarfs} the viscosity tensor is specified
completely from five independent quantities: $K_0$, $K_{0'}$, $K_{00'}$,
$K_1$, $K_2$. Two of these ($K_1$ and $K_2$) were calculated directly in
\citet{our_direct_viscosity}, and for the other three \citet{our_k_dwarfs}
provide perturbative values. In terms of these quantities the time and volume
averaged rate of energy dissipation due to the turbulent flow is given by
\citep[][equation 13]{our_k_dwarfs}:
\begin{eqnarray}
	\dot{\mathcal{E}}_{visc}(\Omega)=\frac{1}{2}\int_0^{L_z} dz \big[
	&&K_1\mathcal{A}_1^2 + K_2\mathcal{A}_2^2 + K_0 \mathcal{A}_0^2 +
	\nonumber\\
	&&+ K_{0'} \mathcal{A}_{0'}^2 + 2K_{00'} \mathcal{A}_0\mathcal{A}_{0'}
	\big],
	\label{eq: viscous power}
\end{eqnarray}
where the $|A_m|^2$ quantities are root mean square shear components defined
by:
\begin{eqnarray}
	\mathcal{A}_0&\equiv&\left<\left(\pard{v_x}{x}+
		\pard{v_y}{y}\right)^2\right>^{1/2}\\
	\mathcal{A}_{0'}&\equiv&\left<\left(
		\parz{v_z} \right)^2\right>^{1/2}\\
	\mathcal{A}_{1}&\equiv&\left<\left| 
		\left(\parz{v_x} + \pard{v_z}{x}\right)
		+ i \left(\parz{v_y} + \pard{v_z}{y}\right)
		\right|^2\right>^{1/2}\\
	\mathcal{A}_{2}&\equiv&\left<\left|
		\left(\pard{v_x}{x}-\pard{v_y}{y}\right)+
		i \left(\pard{v_x}{y} + \pard{v_y}{x}\right)
		\right|^2\right>^{1/2},
	\label{eq: rotation A}
\end{eqnarray}
with the angle brackets denoting a time average.

Equation \ref{eq: viscous power} is simply the volume and time average of the
rate of work done by an anisotropic viscous force on a stratified fluid
subject to some externally imposed shear. The particular form of Eq. \ref{eq:
viscous power} assumes that locally the viscosity tensor is invariant under
rotations around the axis of gravity ($z$), which must be true if stellar
rotation is ignored.

The directly obtained effective viscosity of \citet{our_direct_viscosity} and
the perturbative estimates based on realistic low mass star simulations
\citep{our_k_dwarfs} and on an idealized simulation \citep{our_code} all
produce a linear scaling of the effective viscosity with period, for the
range of periods available to those simulations. As is discussed in these
works the linear scaling is not expected to hold for arbitrarily small
periods, because at such timescales Kolmogorov cascade should be a good
approximation to the flow and in that case \citet{Goodman_Oh_97} show that
the loss of efficiency should be quadratic with period. In addition, the
direct calculations \citep{our_direct_viscosity} show that the effective
viscosity saturates at long periods.

To accommodate these three scalings, we will assume the following form for the
viscosity coefficients:
\begin{eqnarray}
	K_m&=&\min\left[s_m^*\left(\frac{P}{\tau}\right)^2; 
		s_m\frac{P}{\tau}+\Delta_m; s_m\Pi_{max}+\Delta_m\right]\times
		\nonumber\\
		&&\times\frac{1}{3} \rho
		\left<v^2\right>^{1/2} l,
\end{eqnarray}
with:
\begin{equation}
	s_m^*=\frac{s_m}{\Pi_{min}}+
		\frac{\Delta_m}{\Pi_{min}^2},
\end{equation}
where, $l$ is the mixing length, and $\Pi_{min}$ and $\Pi_{max}$ are the
dimensionless periods between which the linear scaling applies, $s_m$ are the
set of slopes with period of each effective viscosity component, and
$\Delta_m$ are the corresponding zero crossings.

This expression has been chosen to be continuous with period ($P$), reproduce
the quadratic scaling we expect to occur at small periods, the linear scaling at
periods comparable to the local convective turnover time ($\tau$) with a
possible offset ($\Delta_m$), and a constant effective viscosity for much
longer periods.

From \citet{our_direct_viscosity}:
\begin{equation}
	s_1=0.084,\quad s_2=0.055.
\end{equation}
Assuming that the scaling between the slopes of these components and the rest
from the \citet{our_k_dwarfs} perturbative calculation holds, we arrive at
the following values for the remaining three $s$ values:
\begin{equation}
	s_{0'}=0.23,\quad s_0=0.07,\quad s_{00'}=0.02
\end{equation}

The offsets, $\Delta_m$, are less well constrained from the
\citet{our_direct_viscosity} and \citet{our_k_dwarfs} simulations,
because on one hand their values from the direct calculation depend on the
method used to extract the effective viscosity and on the other the perturbative
calculation does not predict any offsets so it is not clear of how to find them
for the remaining components. 

For the components not corresponding to radial shear we will assume no offset.
The $\Delta_1$ component is expected to be between the
values derived by the two direct methods of \citet{our_direct_viscosity} 
for a forcing strength of zero. Ignoring the
difference between the weak forcing and a zero forcing case, which according to
figure 13 of \citet{our_direct_viscosity} is likely to be small, we conclude that:
\begin{equation}
	0.023<\Delta_1<0.061
	\label{eq: 1 offset}.
\end{equation}
For $\Delta_{0'}$ we will show results with two different assumptions:
\begin{equation}
	\Delta_{0'}^0=0,\quad\Delta_{0'}^1=2.6\Delta_1,\quad
	\label{eq: 0' offset}
\end{equation}
The last case corresponds to assuming that the offset scales the same way as the
slope between the $m=1$ and $m=0'$ components.

Finally we need to specify $\Pi_{max}$ and $\Pi_{min}$. From
\citet{our_direct_viscosity} we see that $\Pi_{max}=2.4$ is reasonable. The
value of $\Pi_{min}$ is not seen in the simulations, but clearly
$\Pi_{min}<0.3$, so we will consider two cases $\Pi_{min}=0.1$ and
$\Pi_{min}=0.01$.

\section{Tidal Torque}
\label{sec: torque}
An analytical expression for the tidal velocity and the associated tidal torque
was derived with the smallest number of assumptions by \citet{Scharlemann_81,
Scharlemann_82} for a circular orbit and close to synchronous internal rotation.
Here we will include the relevant results and refer to those works for their
derivation.

\citet{Scharlemann_82} equation (17) gives the tidal torque as:
\begin{equation}
	T=\frac{96\pi}{5} r_*^3 n
	\omega_{2*}f_*^2\Lambda_1(\alpha_0-\alpha_{0c}),
\end{equation}
where, $r_*$ is the radius of the star, $f_*$ is the maximum deviation of the
equipotential surfaces from spheres, $n$ is the orbital angular frequency,
$\Lambda_1$ can be expressed through integrals of the viscous force over the
star, $\alpha_0-\alpha_{0c}$ is the departure from synchronous rotation,
and $\omega_{2*}$ and $\alpha_0$ describe the differential rotation assumed to
have the form:
\begin{equation}
	\Omega(r,\theta)=n\omega_{2*}\left(\alpha_0+\sum_l \alpha_l x^l +
	\frac{3}{2} x^2 \sin^2\theta\right).
\end{equation}
The second term in the above expression does not enter in 
$\Lambda_1$ and so will not play any further role.

The value of $\Lambda_1$ is given by \citet{Scharlemann_82} equation (16):
\begin{equation}
	\Lambda_1=I_{r2}+\frac{1}{7}I_{\theta 2} + \frac{20}{7} I_{\phi 2} -
	\frac{25}{28} I_{\phi 4} - \frac{5}{14} I_{\theta 4},
\end{equation}
where the indices, $2$ and $4$, refer to the $P_2^2(\cos\theta)$ and
$P_4^2(\cos\theta)$ associated Legendre polynomial components respectively of
the following integrals of the viscous force ($F$) acting on the tidal velocity
field (\citet{Scharlemann_82} equations (5 c-e)):
\begin{eqnarray}
	I_r(\theta)=\int_{r_b}^{r_*} \frac{r^4 F_r}{g} dr,\\
	I_\theta(\theta)=\int_{r_b}^{r_*} \frac{r^4 \tan\theta F_\theta}{g}dr,\\
	I_\phi(\theta)=\int_{r_b}^{r_*} \frac{r^4 \sin\theta F_\phi}{g} dr.
\end{eqnarray}

\section{Stellar Models}
\label{sec: stellar models}
The viscosity tensor enters in the calculation of $\Lambda_1$ only through its
$8^{\textrm{th}}$ order moments:
\begin{equation}
	E_m=\int_{r_b}^{r_*} r^8 K_m(P,r) dr,
	\label{eq: E8}
\end{equation}
The depth dependence of $K_m$ arises
from the fact that we expect $K_m$ to scale as $\rho lv/3$ ($\rho$, $l$ and $v$
are the density, the mixing length and the convective velocity respectively),
all of which are functions of depth and it also depends of the convective
turnover time, approximated as $l/v$, which also varies with depth.

In order to evaluate the right hand side of equation \ref{eq: E8} we need
$\rho$, $l$ and $v$ in the interior of the star. We will
take these quantities from two stellar models: a $1.4M_\odot$ and a $0.8M_\odot$
main sequence solar metalicity models calculated using the code described in
\citet{Cody_Sasselov_02}.

The reason for choosing these two models is that presently almost all the known
transiting extrasolar planets are around stars that fall within this mass range,
so our results will span the range of currently known systems.

\section{Orbital Evolution}
\label{sec: orbit}

\begin{figure}
	\begin{center}
		\includegraphics[width=0.5\textwidth]{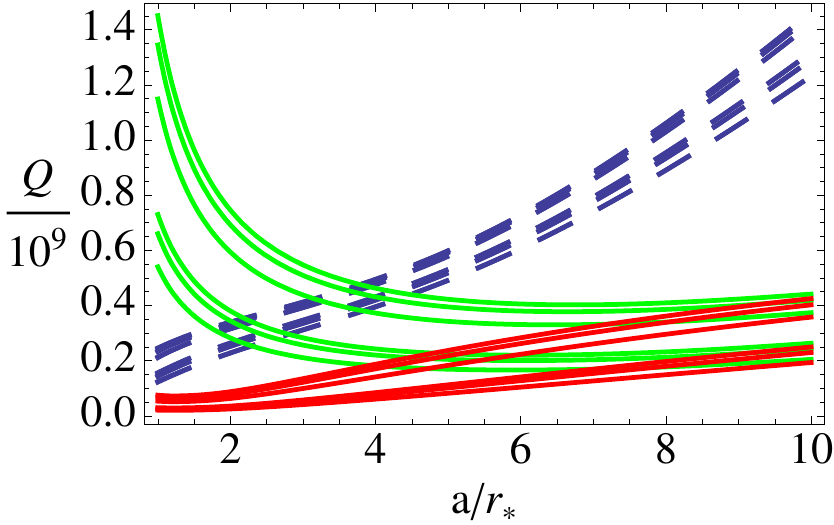}
		\caption{The tidal quality factor corresponding to the
		combined viscosity from \citet{our_direct_viscosity,
		our_k_dwarfs, our_code} for the range of assumptions described in
		Section \ref{sec: viscosity}. The blue dashed curves correspond to
		the $1.4M_\odot$ model, the remaining curves are for the
		$0.8M_\odot$ model: $\Pi_{min}=0.1$ --- green, $\Pi_{min}=0.01$
		--- red}
		\label{fig: Q}
	\end{center}
\end{figure}

From the tidal torque we can calculate the rate at which the orbit of a
circularized and synchronized planet orbiting a slowly rotating star shrinks.
For comparison purposes we converted this rate to an effective $Q_*$ value,
shown in figure \ref{fig: Q}. 

As we can see, for both stellar models and all assumptions of Section
\ref{sec: viscosity}, the dissipation efficiency is much smaller than the
usually assumed values of $10^5<Q_*<10^7$, implied observationally for main
sequence binary stars. As discussed in the introduction we expect that this
is due to the fact that in the case of planets the tides have a frequency
that is too high to resonantly excite inertial waves in the star.

The reason for twice as many curves being visible for the $0.8M_\odot$ case
than for the $1.4M_\odot$ is that the two different assumptions for
$\Pi_{min}$ are indistinguishable for the higher mass model due to the fact
that the convection in the high mass star is much more vigorous and the
orbital period does not get below 0.1 convective turnover times for most of
the convective zone for the range of separations considered. The dissipation
being more efficient for the low mass star is mostly due to the fact that it
is much less centrally concentrated and hence a lot more mass is subject to
the tides than for the high mass case.

\begin{figure}
	\begin{center}
		\includegraphics[width=0.5\textwidth]{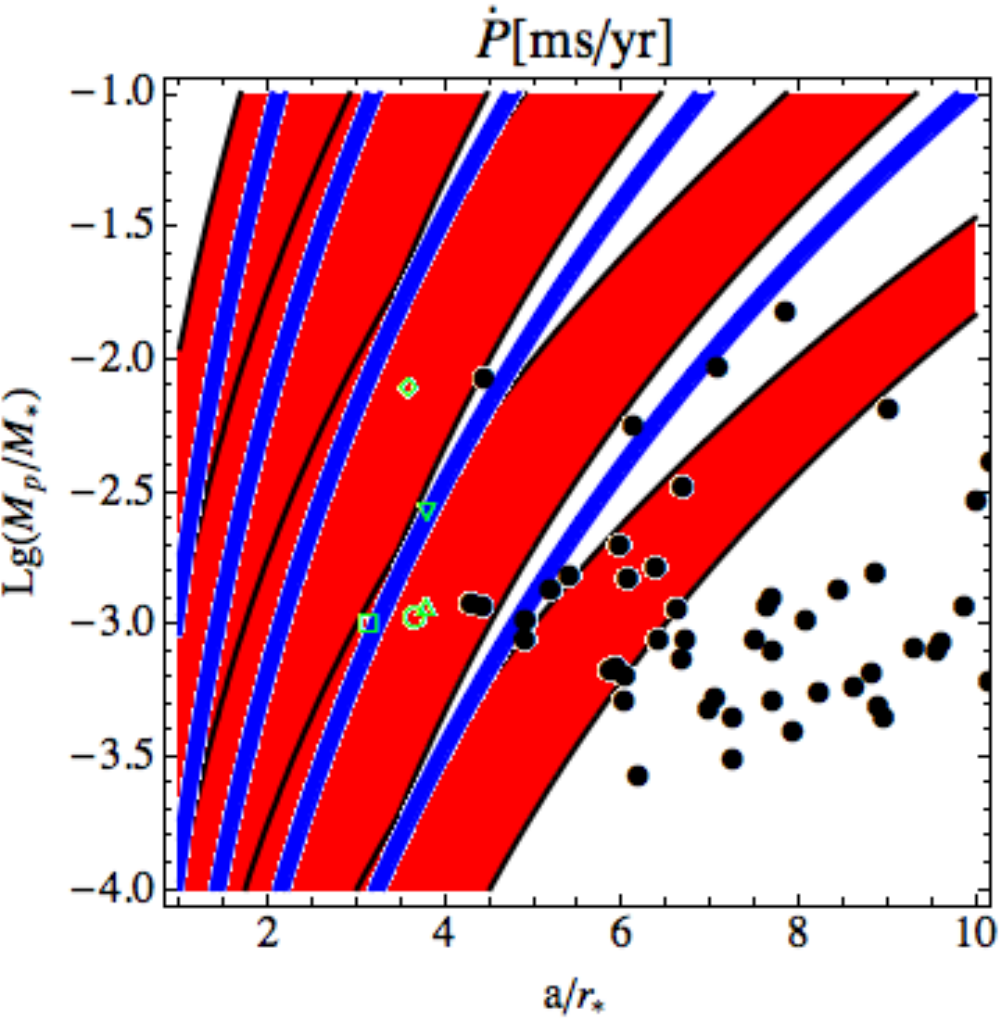}
		\caption{The rate of change of the orbital period of the planet
		in ms/yr. The colored regions correspond to the range of assumptions
		discussed in Section \ref{sec: viscosity}, blue for the $1.4M_\odot$
		model; red for the $0.8M_\odot$ model. The black circles and
		the green symbols  correspond to the currently known transiting
		extrasolar planets that fall in the plotted range. For each stellar
		model, the values of the TTVs for the regions going from left to
		right are 10, 1, 0.1, 0.01 and 0.001 respectively.}
		\label{fig: Pdot}
	\end{center}
\end{figure}

In figure \ref{fig: Pdot} we show the TTVs that will be produced by this
turbulent viscosity. Each colored region corresponds to the range of
viscosity assumptions listed in Section \ref{sec: viscosity}, blue for the
$1.4M_\odot$ model, and red for the  $0.8M_\odot$ model and the red to the
most dissipative. For reference we have added the currently known transiting
extrasolar planets that fall within the plotted range, marking the ones
most affected by tides separately: WASP-12b -- square, WASP-18b -- rhomb,
OGLE-TR-56b -- circle, WASP-19b -- up triangle, WASP-33b -- down triangle. 
As can be seen, for even the closest and most massive extrasolar planets the
rate of period change is less than a ms/yr.

The small TTVs of course are to be expected since planets which decrease
their orbital period quickly will exist only for a short amount of time
(relative to the main sequence life of their parent star) before falling inside
the Roche lobe and being tidally destroyed, making the probability of finding
such planets small. They are an order of magnitude smaller than what was predicted
for OGLE-TR-56b using the older linear prescription for dissipation
\citep{Sasselov_03}.
However, if such close-in giant planets are discovered by $Kepler$, an
extended multi-year mission could achive sufficient precision to constrain
the theoretical work used here.

To address the question of tidal destruction of planets we used the
expression for the tidal torque to calculate the
coupled evolution of the planetary orbit and the stellar spin for systems that
are assumed to start with a non-spinning star and a planet in a circular orbit.
We assume that the planet is synchronized, so we do not consider the tides on
the planet and evolved the system forward in time until the dissipation
drives the planet close enough to its star to be tidally destroyed, thus getting
an estimate of the lifetime of the planet during which it will be observable. 

\begin{figure}
	\begin{center}
		\includegraphics[width=0.5\textwidth]{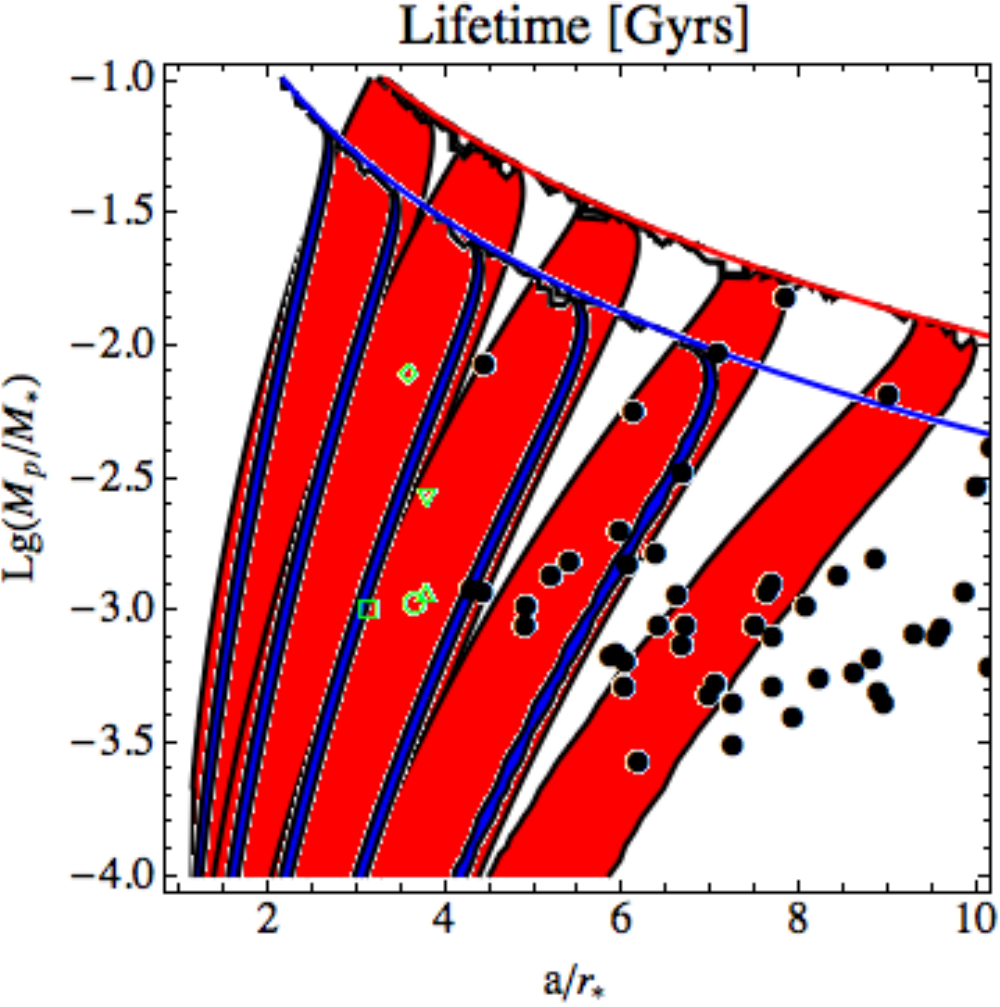}
		\caption{The future lifetimes of single planet systems on a
		circular orbit around an initially non-rotating star for the two
		stellar models considered in the text: $0.8M_\odot$ --- red,
		$1.4M_\odot$ --- blue. Each colored region corresponds to the range
		of assumptions from Section \ref{sec: viscosity}. The black points
		and green symbols correspond to the currently known transiting solar
		planets that fall within the plotted range of masses and
		separations. The remaining lifetimes for each model corresponding to
		the shown regions are 0.01, 0.1, 1, 10 and 100 Gyrs going from left
		to right.}
		\label{fig: lifetimes}
	\end{center}
\end{figure}

These lifetimes for a range of planet masses and separations are shown in
figure \ref{fig: lifetimes} for the two stellar models discussed above. Each
pair of blue curves shows the most and least dissipative effective viscosity
assumptions for the $1.4M_\odot$ model and the red curves show the range for
the $1.4M_\odot$ model. The lines to which the curves for a given model
asymptote in the upper right corner correspond to the critical planet-star
separation beyond which the planetary orbit has sufficient angular momentum
to eventually spin up the star to synchronous rotation thus halting tidal
evolution. Again for reference we have shown the known transiting planets.

As we can see most of the currently observed planets have tidal lifetimes of
many billions of years, and (after considering Wasp-12b, Wasp-18b,
Ogle-TR-56b and WASP-19b individually) we see that no planet has a lifetime
of less than a few hundred million years, the shortest lived being Wasp-18b
with a lifetime ranging between 100 and 160 Myrs for a $1.25M_\odot$ stellar
model \citep[the mass quoted in][]{wasp18_nature}.

\section{Discussion}
\label{sec: discussion}

In this paper we have presented a direct theoretical calculation of the
dissipation efficiency of tides raised on stars by a short period giant
planet. The estimates of this efficiency are based on direct simulations of
the dissipation of externally driven perturbations in a small, but
still significantly stratified, piece of a convective zone.

While due to numerical limitations we are unable to explore the entire
range of the ratio of tidal frequency to local convective turnover time that
occurs in the surface convective zones of low mass stars, we show that the
rate of dissipation of tidal energy averaged over the entire convective zone
is constrained to within a factor of ten. This is a significant improvement
over the current range of three to five orders magnitude that many authors
wishing to calculate the tidal evolution of exoplanet orbits are forced to
consider \citep[e.g.][]{wasp18_nature, Jackson_et_al_09b, Barker_Ogilvie_09,
Levrard_et_al_09, Patzold_et_al_04, Adams_Laughlin_06}. 

Since the usual way to parametrize the dissipation of tides is using the
stellar quality factor $Q_*$, we calculate effective values for this
parameter as a function of the orbital frequency. We find that our estimates
lie towards the upper end (small dissipation) of the usually assumed range
for this parameter -- $10^8<Q_*<10^9$ -- and are significantly larger than the value needed to explain the fact that the maximum period to which solar type binary stellar orbits are circularized increases noticeably over the main sequence lifetime of these systems. 

We argue, however, that this is not necessarily a contradiction, since in tha
case of binary stars synchronization of the stellar rotation to the orbit
happens much faster than the circularization of the orbit. This means that
the tidal frequency is close to the rotation rate of the star, for most
of the time circularization occurs. This opens the possibility that inertial
waves are resonantly excited in the stars, which results in a much enhance
dissipation \citep{Savonije_Papaloizou_97, Dintrans_Ouyed_01, Ogilvie_Lin_04,
Wu_05a, Wu_05b, Papaloizou_Ivanov_05, Ogilvie_Lin_07, Ivanov_Papaloizou_07,
Ogilvie_09}, over what is likely to occur in planet star systems, which
typically have the star rotating much slower than the tides raised by the
planet. 

This idea seems to be confirmed by a number of transiting extrasolar planets
which are found to lie very close to their host star. So close in fact that
were the value of $Q_*$ implied by main sequence binary stars applicable to
star-planet systems, we would conclude that we caught several of those
planets in the last less than 0.1\% of their lifetime
\citep[e.g.]{wasp18_nature, Hebb_et_al_10}, which seems rather unlikely given
the current sample of about a hundred known transiting planets. 

We show that our much smaller dissipation results in future lifetimes for all
the currently known transiting planets comparable to their age (Fig.
\ref{fig: lifetimes}), and is hence consistent with those observations. 

We also calculate the inferred TTVs due to tidal decay of the orbit (Fig.
\ref{fig: Pdot}) and find values of $\lesssim$ 1ms/yr for all currently know
transiting planets. Much smaller than we can hope to observe from the ground
in the near future, but perhaps within reach of an extended Kepler mission if
a very short period massive planet happens to be among its targets, leaving
the possibility to measure $Q_*$ directly.

\bibliography{convective_turbulence}
\bibliographystyle{apj}

\end{document}